\documentclass[namedreferences]{kluwer}

\usepackage[dvips]{graphicx}

\usepackage{makeidx}

\makeindex

\begin{document}
\begin{article}

\begin{opening}
\title{TeV gamma rays from  the Galactic Center}
\subtitle{Direct and indirect links to the massive black hole in Sgr A*}
\author{Felix  \surname{Aharonian} \email{felix.aharonian@mpi-hd.mpg.de}}
\institute{Max-Planck-Institut f\"ur Kernphysik, Saupfercheckweg 1, 69117 Heidelberg, Germany}
\author{Andrii  \surname{Neronov} \email{andrii.neronov@epfl.ch}}
\institute{Ecole Politechnique Federale de Lausanne,  BSP, 1015, Lausanne, Switzerland }
\runningauthor{Aharonian and Neronov}
\runningtitle{TeV radiation from GC}
\begin{ao}\\
KLUWER~ACADEMIC~PUBLISHERS PrePress Department,\\
P.O.~Box~17, 3300~AA~~Dordrecht, The~Netherlands\\
e-mail: TEXHELP@WKAP.NL\\
Fax: +31 78 6392500
\end{ao}

\begin{abstract}
The recent detection of TeV gamma-radiation from the direction 
of the Galactic Center  within several arc-minutes around Sgr~A*  
is the first model-independent  evidence of existence of high energy
particle accelerator(s)  in the central 10 pc region of our Galaxy.
This is an extraordinary site that harbours many remarkable objects
with  the compact radio source Sgr~A*  - a hypothetical super-massive 
black hole (SMBH) - in the dynamical center of the Galaxy.  Here we explore  the
 possible {\em direct} and {\em indirect} links of the reported TeV
  emission to the SMBH.  We show that at
  least three $\gamma$-ray production scenarios that take place close
  to the event horizon of the SMBH can explain the reported TeV
  fluxes. An alternative (or additional) channel of TeV radiation is
  related to the run-away protons accelerated in Sgr~A*.
  Quasi-continuous injection of relativistic protons into the
  surrounding dense gas environment initiates detectable high energy
  gamma-ray emission.  The absolute flux and the energy spectrum of
  this radiation component strongly depend on the history of particle
  injection and the character of diffusion of protons during the last
  $10^5$ years.  For a reasonable combination of a few model
  parameters, one can explain the detected gamma-ray flux solely by
  this diffuse component.
\end{abstract}

\end{opening}

\section{Introduction}

The recent detection of TeV gamma-ray emission from the direction of
the Galactic Center by three independent groups, CANGAROO \cite{tsuchiya2004},
Whipple \cite{kosack2004} and HESS \cite{aharonian2004}, is a
remarkable result which will  have a strong impact on our
understanding of high energy processes in the central region of the
Galaxy.  The localisation of the TeV signal by HESS within a few
angular minutes indicates that the gamma-ray source(s) is (are)  located in
the central $\leq 10$ pc region.  Among the possible sites of
production of TeV gamma-rays are Sgr A*, the young supernova 
remnant Sgr A East,  the Dark Matter Halo,  and finally the whole 
diffuse 10 pc region. It is possible that some of these potential gamma-ray
production sites  comparably contribute to the observed TeV flux.
Moreover, the same source could be responsible for two different
components of radiation: (i) the {\em direct} gamma-ray component
produced {\em inside} the particle accelerator, and (ii) the {\em
  indirect} component produced by runaway protons which are accelerated in the
same source, but later injected into the surrounding dense gas environment.
 
Below  we discuss these two components of TeV radiation  in the context of 
acceleration of  particles in the proximity of the SMBH   and their radiation both 
inside and outside of Sgr A*.    

\section {Production of high energy gamma-rays  in Sgr A*}

The temporal and spectral features of Sgr A* are unusual and, as
a whole,  different from other compact galactic and
extragalactic sources containing black holes.  This concerns, first of
all, the extraordinary low luminosity of Sgr A*.  In addition the other
important astrophysical implications, the low luminosity of Sgr A* has a dramatic
effect on the visibility of the source in gamma-rays \cite{apj-paper}.

\subsection{Transparency of Sgr A*  for high energy gamma-rays}

\begin{figure}
\begin{center}
\resizebox{0.9\hsize}{!}{\includegraphics[angle=0]{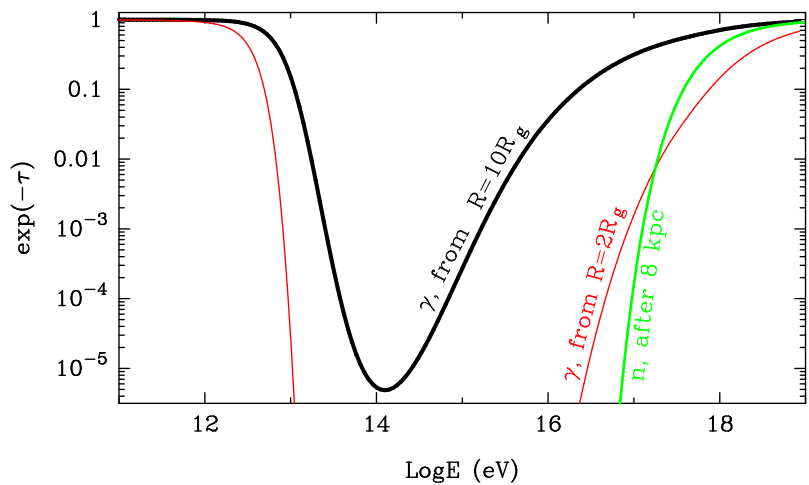}}
\caption{Attenuation of gamma-rays in Sgr A* due to internal
  photon-photon pair production dominated by interactions of high
  energy gamma-rays with radiation of the compact infrared source.
  Two solid curves
  marked ``$\gamma$'' are calculated assuming that the infrared
  emission of Sgr A* is produced within $10R_g$ and $2R_g$ around the
  central black hole of mass $3 \times 10^6 M_\odot$.  
  The curve marked ``n'' shows attenuation of
  the neutron flux $\exp{(-d/\Lambda)}$, where $\Lambda \approx 10
  (E/10^{18}$ eV$) \ \rm kpc$ is the decay mean free path of a neutron
  of energy $E$, and $d=8 \rm kpc$ is the distance to the Galactic
  Center.}
\label{fig:opt_depth}
\end{center}
\end{figure}

Because of internal photon-photon pair production, 
the high energy gamma-ray emission of these objects (both of
stellar mass and super-massive BHs) is generally suppressed, and
consequently the unique information on possible particle acceleration
processes near the event horizon of the BH is essentially lost.
But  this is not the case of the super-massive BH located at
the dynamical center of our Galaxy (Sgr~A*), which thanks to its
extraordinary low bolometric luminosity ($ \leq 10^{-8} L_{\rm Edd}$)
is transparent for  very high energy gamma-rays. 
It  is seen from Fig. \ref{fig:opt_depth} that indeed up to 10
TeV the source is transparent for gamma-rays even if one assumes that
gamma-rays are produced  within $2 R_{\rm g}$.
Note that the decrease of the $\gamma \gamma \rightarrow e^+e^-$  cross-section well above
the pair production threshold makes the source again transparent at $E \sim 10^{18} \ \rm eV$

High energy gamma-rays from compact regions close to the event horizon
of the SMBH  can be produced in various ways due to acceleration
of protons and/or electrons and their interactions with ambient
magnetic and radiation fields, as well as with the thermal plasma.

\subsection{Synchrotron and curvature radiation of protons}

Synchrotron and curvature radiation of protons  are important processes 
in the so-called extreme accelerators \cite{aharonianetal2002} where 
particles are accelerated at the maximum possible rate, $\dot{E}=eB$.
However, even for an ``ideal'' combination of parameters allowing the
most favourable acceleration/cooling regime, the characteristic energy of
synchrotron radiation is limited by
$\epsilon_{\rm max}=(9/4) \alpha_f^{-1}  m_pc^2 \simeq 0.3 \mbox{ TeV}$   
\cite{aharonian2000}.
This implies that the proton-synchrotron radiation cannot explain the
flux observed from the direction of GC up to several TeV,
unless the radiation takes place in a source moving
towards the observer with bulk motion Lorentz factor  $\Gamma \geq 10$.
Another possibility could be if the {\em proton-acceleration} and  
{\em synchrotron gamma-ray production} regions are {\em separated},  e.g. 
when protons are accelerated in a
regular B-field while moving along field lines, and later
are injected into a region of chaotic magnetic field. 

In this scenario we should expect $\gamma$-rays also from the regular B-field region - 
due to the proton curvature radiation.  The contribution of latter  
in the high energy radiation of SMBHs  could be quite significant \cite{levinson2000}.
The curvature radiation
of protons can extend to  
$\epsilon_{\rm max} =3 E_p^3 / 2 m^3R \simeq 0.2 (B / 10^4 \ \rm )^{3/4} \mbox{TeV}$  
(hereafter all estimates correspond to the black-hole mass 
$3 \times 10^6 M_\odot$).
Formally, this equation allows extension of the spectrum  to 10 TeV, provided that the 
magnetic field exceeds $B \simeq 10^6$ G.  However,  such a strong
field would make  the source  opaque for TeV gamma rays
\cite{apj-paper}.

\subsubsection{Photo-meson interactions}

The protons accelerated in the region close to the event horizon of SMBH to
energies  $E \sim 10^{18} \ \rm eV$, start to interact with
soft photons of the compact infrared source located at 
$\sim 10 R_{\rm  g}$.  Despite the low luminosity of the source, the density of
infrared photons appears sufficiently high for reasonably effective photo-meson
interactions.  Indeed, the mean free path of protons through the
photon field is estimated $\Lambda_{p\gamma}\sim  (\sigma_{p \gamma} f n_{\rm ph})^{-1}  
\simeq 10^{15} (R_{\rm IR} / 10^{13}\mbox{ cm})^2 \mbox{ cm}$.  
This means that approximately  $R/\Lambda_{p\gamma} \sim 0.01$ fraction of the energy 
of protons is converted into secondary particles (neutrinos, photons, electrons).

\begin{figure}
\begin{center}
\resizebox{0.9\hsize}{!}{\includegraphics[angle=0]{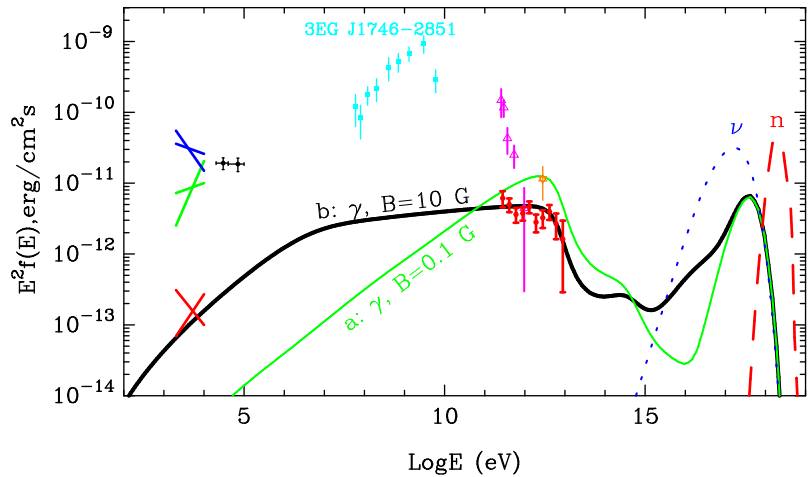}}
\caption{Broad-band Spectral Energy Distribution (SED)  of gamma-rays (solid lines), neutrons
  (dash line) and neutrinos (dots) from Sgr A* due to interactions of
  ultra high energy protons with ambient photon and magnetic fields.
  Protons accelerated to energies
  $10^{18}$ eV in the regular magnetic field close to the
  gravitational radius $R \sim R_{\rm g}$, propagate through the
  infrared emission region of size $R=10 R_{\rm g}$.  The calculations
  correspond to two assumptions for the strength of the magnetic field
  in the region of the infrared emission: $B=0.1$ G (a) and $B=10$ G
  (b).  The experimental points in X-ray and gamma-ray bands 
are from the compilation of  Aharonian and Neronov (2005).  }
\label{fig:pgamma}
\end{center}
\end{figure}

While neutrinos and neutrons, as well as gamma-rays of energies below
$10^{12}$ eV escape freely the
emission region, gamma-rays above  $10^{12}$
as well as secondary electrons from $\pi^\pm$-decays
effectively interact with the ambient photon and magnetic fields, and
thus initiate IC and/or (depending on the strength of the B-field)
synchrotron cascades.  The cascade development stops when the typical
energy of $\gamma$ rays is dragged to 1 TeV. 
Gamma-rays produced in this way can explain the observed TeV flux, if
the acceleration power of $10^{18}$ eV protons is about $10^{37} \ \rm
erg/s$.  The energy spectra of gamma-rays produced in this scenario
are shown in Fig.~\ref{fig:pgamma}.

%

\subsubsection{Proton-proton scenario}
Acceleration of protons to extremely high energies, $E \sim 10^{18}$
eV, is a key condition  for  effective  photo-meson
interactions.  This model requires  existence of strong magnetic field, $B
\geq 10^4$ G, in the compact region limited by a few gravitational
radii.  If the field close to the black hole is weaker,
the  interactions of protons with  the ambient thermal gas 
become the main source of production of gamma-rays and
electrons of ``hadronic'' origin. 

\begin{figure}
\begin{center}
\resizebox{0.9\hsize}{!}{\includegraphics[angle=0]{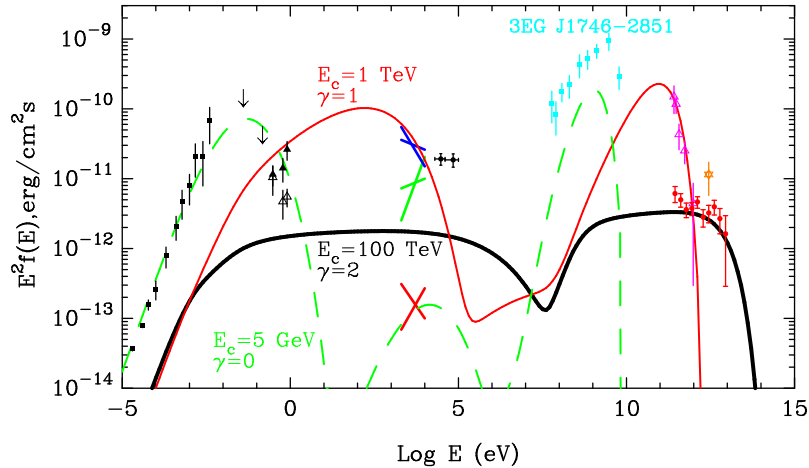}}
\caption{SED  of the broad-band electromagnetic
  radiation initiated by $p-p$ interactions in the accretion disk
  (from Aharonian and Neronov 2005).  It is assumed that the
  accelerated protons with spectrum $E^{-\Gamma} \exp{(-E/E_c)}$ are
  injected into the thermal plasma of density $10^8 \ \rm cm^{-3}$,
  and together with the accretion flow cross the region of the size
  $R\approx 10R_g$ cm and fall under the black hole horizon after
  $10^4$ s.  The following parameters have been assumed: (1) heavy
  solid curve: $\Gamma=2$, high energy exponential cut-off at
  $E_c=100$ TeV, total acceleration rate $L_{\rm p}=5 \times 10^{38} \ 
  \rm erg/s$; (2) thin solid curve: $\Gamma=1$, $E_c=1$ TeV, $L_{\rm
    p}=10^{40} \ \rm erg/s$; (3) dashed curve: narrow
  ($\Gamma=0,E_c=5$~GeV) distribution of protons, $L_{\rm p}=10^{40}$
  erg/s. For all three cases the magnetic field is assumed to be
  $B=10$ G.  }
\label{fig:pp}
\end{center}
\end{figure}

Protons can be accelerated to TeV energies also in the accretion disk,
e.g.  through strong shocks developed in the accretion flow. The
efficiency of gamma-ray production in this case is determined by the
ratio of accretion time $R/v_{\rm r} \sim 10^3 - 10^4$ s (depending on
the accretion regime) to the 
{\em p-p}  cooling time, $t_{\rm pp} \simeq 1.5
\times 10^{7} (n/10^8\mbox{ cm}^{-3})^{-1}  \mbox{ s}$. 
For any reasonable assumption concerning the density of the ambient
thermal plasma and the accretion regime, the acceleration power  of high energy
protons should exceed $L_{\rm p} \approx 10^{39} \ \rm erg/s$ in order
to provide detectable fluxes of TeV gamma-rays  (see Fig. \ref{fig:pp}).  

%

\subsection{Curvature Radiation - Inverse Compton  (CRIC) model}

The models of gamma-ray emission associated with accelerated
protons provide rather modest efficiencies  of conversion of the energy of 
accelerated protons to  gamma-rays.  The radiative energy loss rate of electrons
is much higher, and therefore the models associated with accelerated
electrons provide more economic ways of production of high energy
gamma-rays.  Obviously, these electrons should be accelerated to
at least $E_{\rm max} \sim 10 \ \rm TeV$.  This immediately constrains the strength of the
chaotic component of the magnetic field; even under an  extreme 
assumption that the acceleration proceeds at   the maximum possible rate, 
$(dE/dt)_{\rm acc}  \simeq eB$, one gets
$B \leq 10 (E_{\rm max}/10 \rm TeV)^{-2} \ \rm G$.

The requirement of particle acceleration at the maximum rate imposes
strong restrictions on the geometry of magnetic field and possible
acceleration mechanisms.  In this regard, acceleration in ordered
electric and magnetic fields, e.g.  by the rotation-induced electric,  
provides maximum energy gain. Moreover, in
the ordered field the energy dissipation of electrons is reduced to
the curvature radiation loses;  this  allows  acceleration of electrons up to  
$E_{\rm e, max} \simeq 10^{14} (B /10 \rm G)^{1/4} \mbox{ eV}$.

The curvature radiation peaks at 
$\epsilon_{\rm curv} \simeq 2 \times 10^8 (E_{\rm e} /10^{14} \rm eV)^3 \mbox{eV}$.  
The Compton scattering of same electrons
leads to the second peak at much higher energies, $E_\gamma \sim E_{\rm e}
\sim 10^{14} \ \rm eV$ (because the  scattering proceeds in the
Klein-Nishina limit).  However, because of 
interactions  with infrared photons,  gamma-rays of energy exceeding   
10 TeV can not freely escape the source. Synchrotron radiation and Compton
scattering of the secondary (pair-produced) electrons lead  to re-distribution of the initial
gamma-ray spectrum. 

\begin{figure}
\begin{center}
\resizebox{0.9\hsize}{!}{\includegraphics[angle=0]{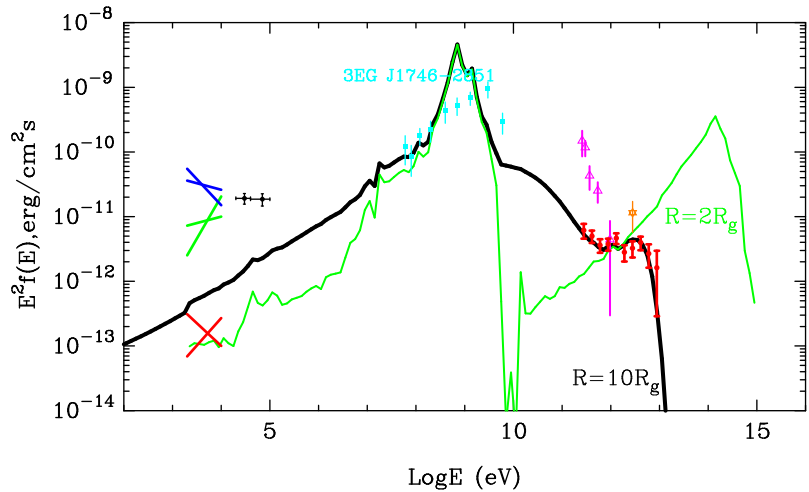}}
\caption{Broad-band SED  of radiation
  produced by electrons within the {\em CRIC} model (from Aharonian
  and Neronov 2005).  Thin solid curve  - gamma-ray production spectrum
  formed as superposition of the Curvature and inverse Compton
  emission components which accompany electron acceleration by the
  rotation-induced electric field within $R=2 R_{\rm g}$; heavy solid 
  curve - the spectrum of gamma-rays modified after the passage
  through the infrared source of size $R=10 R_{\rm g}$. The strength
  of the regular magnetic field in the electron acceleration region is
  assumed $B=10$ G.  The strength of the random magnetic field in the
  region of infrared emission is assumed $B=30$ G.  }
\label{fig:sic}
\end{center}
\end{figure}

We call this scenario of production of {\em C}urvature {\em R}adiation 
and {\em IC}  photons by electrons accelerated in regular magnetic/electric 
fields as {\em CRIC} model.  Quantitative calculations of high energy radiation
within  framework of this model require  a ``self-consistent''
approach which should take into account
the radiation reaction force.  An example of such self-consistent computation is shown in
Fig.  \ref{fig:sic}.  

\section{Diffuse gamma radiation of runaway protons}

A significant fraction of protons accelerated near the black hole may
escape the source and enter the surrounding dense gas environment. The
interactions of these runaway protons with the interstellar medium
lead to production of gamma-rays the luminosity of which could
exceed the gamma-ray luminosity of the central engine.
 
The flux of this radiation component depends not only on the
density of the ambient gas and the injection rate of protons, but also 
on the speed of their propagation in the interstellar
medium. The TeV radiation  detected by HESS is localised  within several angular
minutes. This implies that, for the distance to the Galactic Center 
$d \simeq 8 \ \rm kpc$, the linear size of production region of
gamma-rays can be as large as 10 pc.  The density of the diffuse
interstellar gas in this region is very high, $n \simeq 10^{3} \ \rm
cm^{-3}$.  The lifetime of protons in this dense  environment against
nuclear {\it  p-p}  interactions is $t_{\rm pp} \approx 5 \times 10^4$ yr. Thus, in
the case of absolute confinement of protons, the gamma-ray luminosity
after $\sim 10^5$ years of continuous injection of protons achieves
its maximum (``saturated'') level, $L_\gamma=\eta \dot{W}_{\rm p}$, with
an  efficiency  $\eta \simeq 1/3$  (the fraction of energy released in neutral 
$\pi$-mesons).  However, the confinement time in the 10 pc
region is rather  limited; even in the Bohm diffusion regime  the
escape time of protons  from this region 
is $t_{\rm esc} \sim  R^2/r_g c \simeq
3 \times 10^5 (E/100 \ \rm TeV)^{-1}(B/ 100 \ \rm \mu G)$ yr.  Thus
for any realistic diffusion coefficient $D(E)$,  TeV
protons   leave the region before they loose their energy in  {\it p-p}
interactions. This not only reduces the gamma ray production
efficiency but also, in the case of energy-dependent diffusion, 
modifies  the energy distribution of protons established 
within the 10 pc region.

\begin{figure}
\begin{center}
\resizebox{0.75\hsize}{!}{\includegraphics[angle=0]{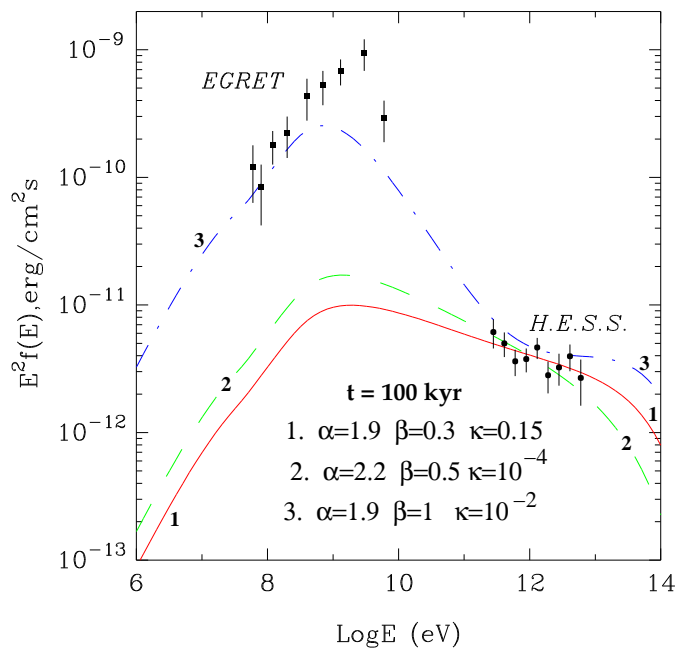}}
\caption{Energy spectra of diffuse gamma rays  expected from the central 10 pc region 
of the Galactic Center within different assumptions concerning the diffusion coefficient 
and the proton injection spectrum (see the text).
}
\label{fig:dif1}
\end{center}
\end{figure}

The impact  of the energy-dependent diffusion on the resulting radiation 
spectra is demonstrated in Fig.\ref{fig:dif1}.  It
is assumed that during $10^5$ years protons are injected  (quasi)continuously into
the interstellar medium of density $n=10^{3} \ \rm cm^{-3}$.  The
initial  spectrum of protons is assumed in the form 
of power-law with an exponential cutoff, 
$Q(E)=Q_0 E^{-\alpha} \exp{(-E/E_0)}$. The
cutoff energy is fixed at $E_0=10^{15} \ \rm eV$, which is an obvious condition 
for effective production of gamma-rays to at least 10 TeV. 
The choice of the power-law index depends on the  assumed diffusion coefficient, if
one intends to explain the energy spectrum of gamma-rays
detected by HESS, $J(E)=(2.5 \pm 0.21) \times 10^{-12} E^{-\Gamma} \ 
\rm ph/cm^2 s \ TeV$ with $\Gamma=2.21 \pm 0.09$ (Aharonian et al.
2004).

The diffusion coefficient is assumed in the following form:  $D(E)=10^{28} (E/1 \ \rm
GeV)^\beta \kappa \ \rm cm^2/s$.  The values of  $\kappa \sim 1$ and
$\beta \sim 0.5$ correspond to the CR diffusion in the galactic disk.
Of course, in the central region of the Galaxy  one 
may  expect significant deviation from the character of particle diffusion in 
``ordinary''  parts  of the  galactic disk. In the case of effective confinement of
protons, e.g. with $\kappa=10^{-4}$ and $\beta=0.5$, the escape time
of multi-TeV protons is comparable with the characteristic time of
{\em p-p} interactions.  This prevents  strong modification  of the initial 
proton spectrum. Therefore, the injection spectrum of protons with
power-law index $\alpha =2.2$ fits quite well the observed TeV
spectrum (curve 2 of Fig.\ref{fig:dif1}).  Because of  the effective
confinement, the required injection rate of protons is rather modest,
$\dot{W}_{\rm p}=7 \times 10^{36} \ \rm erg/s$.

In the case of faster diffusion, the spectral index of injection is
determined as  $\alpha \simeq \Gamma - \beta$.  For
example, for $\kappa=0.15$ and $\beta=0.3$ (a diffusion regime which
corresponds  to the Kolmogorov type turbulence), the $\gamma$-ray
observations are well  explained  assuming the following  parameters   
$\alpha=1.9$ and $\dot{W}_{\rm p}=7.5 \times 10^{37} \ \rm erg/s$ (curve 1 in
Fig.\ref{fig:dif1}).

Finally, one should note that the injection spectrum remains unchanged
with very fast escape. For example, for $\kappa=10^{-2}$ and $\beta=1$
(an energy-dependence close to the one expected in the Bohm diffusion,
but with much larger diffusion coefficient), the particles propagate
in the diffusion regime until  energies of several TeV.  At higher energies 
they escape the source almost rectilinearly on timescales 
$R/c \sim 30$ years.  This implies that while at low energies the
protons  inside the 10 pc region suffer significant spectral deformation, 
$\alpha^\prime \rightarrow \alpha + \beta$, at very high energies the shape 
of the  initial  spectrum is essentially recovered. Such an interesting modulation  
of the proton spectrum is reflected  in the resulting 
gamma-ray spectrum (curve 3 in Fig.\ref{fig:dif1}).  In this case the requirement 
to the injection power of protons is  higher
than in the previous cases, $\dot{W}_{\rm p}=10^{39} \ \rm erg/s$.
 
The $\pi^0$-decay radiation of protons is always  accompanied by
synchrotron radiation of secondary electrons - the products of charged
$\pi$-mesons.  In the case of extension of the proton spectrum to
$10^{15} \ \rm eV$, and for the magnetic field exceeding $100 \ 
\mu \rm G$, the spectral energy distribution (SED) of synchrotron
radiation of secondary electrons peaks in the X-ray 
domain.  This seems an attractive  mechanism for
explanation of the diffuse X-ray emission of the Galactic Center,
given the serious problems of interpretation of this radiation within
the ``standard''  (thermal and nonthermal) models \cite{diffuse-chandra}.
However, for a relatively flat SED of TeV gamma-rays (like the one observed by
HESS) the energy flux of X-rays is always less, by a factor of 3 to
10, than the energy flux of $\gamma$-rays. Therefore for the spectra
shown  in Fig. \ref{fig:dif1} only a small fraction (10 per cent or
so) of the observed X-ray flux can be contributed by secondary electrons.

\begin{figure}
\begin{center}
\resizebox{1.0\hsize}{!}{\includegraphics[angle=0]{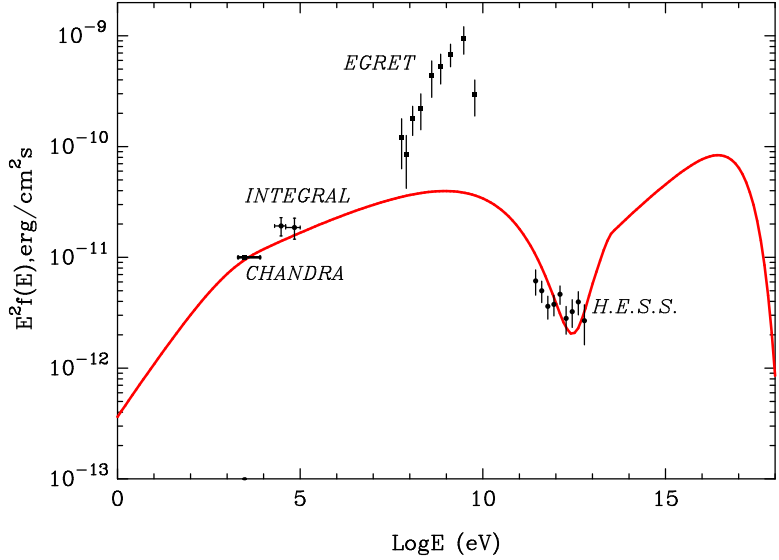}}
\caption{The broad-band spectrum of radiation initiated 
by interactions of ultra-high energy protons in the central 10 pc region. 
The injection spectrum is assumed in the form of power-law with 
$\alpha=1.5$ and low and high energy cutoffs at 
$E_1=2 \times 10^{14} \ \rm eV$ and $E_1=1.5 \times 10^{18} \ \rm eV$.
The proton injection rate  is assumed $\dot{D}=1.5 \times 10^{38} \ \rm erg/s$
to match the reported X-ray (Chandra - Muno et al.  2004, INTEGRAL - Belanger et al.  2004) 
and TeV fluxes (Aharonian et al., 2004).  
The diffusion coefficient is characterised by $\beta=0.3$ 
and $\kappa=10^{-3}$. The number density of the gas  and magnetic field 
in  the central 10 pc region are assumed
$n=10^{3} \ \rm cm^{-3}$ and   $B=1  \ \rm mG$, respectively.
The effect of spectral modulation due  interactions with the 2.7~K CMBR
is not shown. Note that  for the distance to the GC the absorption becomes  noticeable 
($\sim e^{-1} \approx 1/3$) only at energies around $10^{15} \ \rm eV$.  
}
\label{fig:dif2}
\end{center}
\end{figure}

Formally, the flux of the secondary-electron-synchrotron  component 
can be significantly increased,  without getting in conflict with the 
TeV data, if one assumes very hard
spectrum of protons, e.g.  with power-law spectral index $\alpha \leq
1.5$. Such a spectrum cannot explain the TeV data.  However, assuming
that the spectrum of protons extends to $E_0 \sim 10^{18} \ \rm eV$, a
new TeV gamma-component, due to the synchrotron radiation
of the secondary ultrahigh energy electrons,  could dominate over 
the $\pi^0$-decay component.  The possibility of producing high fluxes
of X-rays and gamma rays through the synchrotron radiation of
secondary electrons  is demonstrated in Fig. \ref{fig:dif2}. Note that
although the parent protons interact with the ambient gas throughout
the central 10 pc region of the Galactic Center (thus  the radiation
can be dubbed as diffuse),  the observer would detect a
point-source like TeV signal centered on Sgr A*. The
reason is that the ultra high energy protons 
injected into the surrounding  medium propagate  radially (like photons)  
without significant deflection in the interstellar magnetic field. Since the characteristic 
energy of parent  protons, which are responsible for the secondary synchrotron X-rays,  
is smaller by a factor of (TeV/keV)$^{1/2} \sim 10^4-10^5$, the angular size of X-rays 
should  be larger.  This scenario predicts  
a tendency of decrease of the angular size of  the X-ray signal with 
increase of the photon  energy. This hypothesis
can be inspected also by detection  of direct gamma-rays  from 
photomeson interactions up to $\sim 10^{17} \ \rm eV$.  Above several TeV 
this radiation  component dominates  over the secondary  synchrotron radiation.   

\section{Summary}

The origin  of TeV  radiation  detected from the direction
of the Galactic Center  is not yet established. The 3-arcmin  upper limit on
the  source size,  assuming Gaussian distribution of the source brightness 
(Aharonian et al. 2004), implies that  several objects, in  particular 
the Dark Matter Halo, the young SNR Sgr A East, the central compact source Sgr A*,  
as well as the entire central diffuse  region filled by dense molecular clouds and cosmic rays,   
are  likely candidates for TeV emission within  the central 10 pc region. 
In fact, each  of these sources may contribute significantly into the observed 
gamma-ray flux.  In this paper we 
present the results of our study  of several gamma-ray production scenarios 
with both direct and indirect links to the massive black hole  in  the Galactic Center, Sgr A*.  

The results shown  in Figs.~\ref{fig:pgamma}-\ref{fig:sic} demonstrate 
that at least three gamma-radiation scenarios,  which  take place in 
vicinity of the  massive black hole, can explain the TeV
observations without apparent conflicts with observations of Sgr A* at
lower frequencies.  The hadronic models based on photo-meson or {\it pp}
interactions cannot provide efficiency higher  than 0.1 per cent,
therefore they require an acceleration rate of protons of about  $10^{39} \ \rm erg/s$
or larger.  Although this  is
larger than the total electromagnetic luminosity of Sgr
A*,  it is still acceptable for a black hole of mass $3 \times 10^6
M_\odot$. On the other hand, the electronic model {\em CRIC},  based on
the curvature and synchrotron radiation channels,  allows an  economic
way of conversion of energy of electrons to gamma-rays.

The {\em pp}  hadronic model predicts X-ray fluxes comparable with 
X-ray observations  in high state, but an order of  magnitude
higher that in the quiescent state. The photomeson  and {\em CRIC}  models
predict an order of magnitude lower X-ray fluxes. On the other hand,
the photomeson  model predicts detectable fluxes of
ultrahigh energy gamma-rays and neutrons.
The  {\em CRIC} model predicts high gamma-ray fluxes at MeV/GeV energies
which should be easily detected by GLAST.  

The gamma-radiation
in all three models is generally expected to be variable on timescales as short as 1 hour.
Therefore the detection of a variable component of radiation on such short timescales
would be a strong  argument in favour of gamma-ray production near the massive black hole. 
On the other,  although the lack of variations of the  TeV flux  could be naturally interpreted 
as  gamma-ray production in extended regions,  it cannot be used as 
a  decisive argument against the  black-hole origin of TeV emission.  

In addition to the gamma-rays emitted in  compact regions in the vicinity of the massive  
black hole,  one should expect also a diffuse (extended) component of radiation associated with  
interactions of the runaway protons  with the surrounding dense interstellar gas.
The relative contribution of this component to the total TeV flux, as well as the  spectral and 
angular characteristics of $\pi^0$-decay gamma-rays significantly depend on the rate of 
injection of protons  by Sgr A* into the interstellar medium, as well as on the regime of  
(energy-dependent) diffusion of  protons in the dense central  region. For certain combinations    
of principal   model parameters  the diffuse gamma-rays from {\em pp} interactions can satisfactorily 
explain  both the absolute flux and energy spectrum of TeV radiation reported by HESS. Although 
this component can be  extended  well beyond the Galactic Center region, 
because of enhanced gas density  in the central  10 pc region   one  should expect    
a bright  gamma-ray core within several  arcminutes around Sgr A*.  In the case of extension of 
the spectrum of protons to $E \simeq  10^{18}$ eV,  one may  expect 
another diffuse component related to the synchrotron radiation of ultrahigh 
energy secondary electrons (from $\pi^\pm$-decays). Since the gyroradius of                                      
parent protons is comparable or exceeds 10 pc, the observer would detect this TeV
component as  a point-like source.

\label{lastpage}
\end{article}
\end{document}